# An error-mitigated quantum annealing solution for the weighted Max-Cut problem on a cubic lattice

**Y. S. Yang[1,*], P. Tyson[1,2] and A. B. Murphy[3]**

[1] CSIRO, Locked Bag 10, Clayton South, Victoria 3169, Australia
[2] Present address: Tyson Computing Pty Ltd, Burwood, Victoria 3130, Australia
[3] CSIRO, PO Box 218, Lindfield, NSW 2070, Australia

*Author to whom any correspondence should be addressed.

E-mail: sam.yang@csiro.au

## Abstract

The weighted Max-Cut problem is an NP-hard problem with application implications. It is investigated on a cubic lattice with $11^3$ nodes and mixed-signed random edge weights. For a fixed upper bound on edge weights, it has been demonstrated that the computational difficulty increases as the lower bound on edge weights becomes more negative. The solution time for the problem using a novel error-mitigated quantum annealing approach is compared with standard D-Wave quantum annealing (QA) and BQM hybrid solvers, as well as various classical solvers. For the QPU-embeddable weighted Max-Cut instances with mixed-signed edge weights, it has been quantitatively demonstrated that the SEMO (spin-error mitigation for optimisation) error-mitigated quantum annealing achieved substantially shorter time-to-solution than standard D-Wave QA, D-Wave BQM, simulated annealing and Tabu search baselines. The error-mitigated quantum annealing approach presented in this article potentially elevates the efficiency and application scope of quantum annealing and would be applicable in solving other discrete optimisation problems that can be formulated as QUBO or Ising instances. The promising solution time advantage would be particularly impactful for time-critical optimisation applications.

## 1. Introduction

The Max-Cut problem is a classic optimisation problem in graph theory [1]. Given an undirected graph, the goal is to find a way to divide the graph's nodes into two disjoint sets such that the number of edges between the two sets is as large as possible. In other words, one wants to "cut" the graph in a way that maximises the number of edges that go across the cut. The Max-Cut problem can be solved efficiently and approximately using a heuristic Quantum Approximate Optimization Algorithm (QAOA) [2,3]. The weighted Max-Cut problem on a 3D cubic lattice is a generalisation of the standard Max-Cut problem, in which the nodes form a cubic lattice and neighbouring nodes are linked by edges with varying weight numerical values, which may represent correlation, connection, cost, distance or importance for various applications. The goal of weighted Max-Cut is to divide the nodes of the graph into two disjoint sets so that the sum of the weights of the edges crossing the cut is as large as possible.

In a physics context, the weighted Max-Cut problem on a cubic lattice is equivalent to finding the ground state of an Ising spin glass model with interaction weights [4]. It is relevant to various applications such as community detection in complex networks [5] and energy network optimization [6].

The weighted Max-Cut is a well-known NP-hard problem, meaning there is no known polynomial-time algorithm to solve it exactly for large graphs. A foundational and widely cited reference for the general Max-Cut problem is the Goemans–Williamson (GW) algorithm [7,8], which provides a strong approximation guarantee using semidefinite programming (SDP). However, the GW algorithm is ineffective in solving the weighted Max-Cut problem with mixed-signed edge weights. QAOA solution of the weighted Max-Cut problem was also investigated [9].

Quantum annealers are approximate realisations of adiabatic quantum computing, where solving an optimisation problem is formulated as finding the ground state of a quantum system when the system evolves from a known simple system to a system corresponding to the underlying optimisation problem [10]. Quantum annealing (QA) has emerged as a promising, efficient computation technology for solving complex real-world problems such as scheduling and logistics [11], power network optimisation [12] and advanced materials development [13,14]. Lattice-style simulations in material science often require a large number of lattice nodes far beyond the number of qubits in a quantum processing unit (QPU). Lattice-style spin-glass problems larger than annealer QPU embedding have been attempted [15]. Despite potential advantage over classical computing with QA to obtain global optimal solutions for certain discrete optimisation problems, the accuracy and efficiency are hampered by qubit errors [16], limiting the scope of QA for practical applications.

A significant amount of effort has been devoted by the R&D community to address qubit errors in quantum computing. One mainstream approach is to compute with error-corrected logical qubits [17]. This approach is limited because it requires many physical qubits to construct an error-corrected logical qubit. It substantially reduces the ability of quantum computers to solve large problems that require many qubits. Additionally, the pathway for using error-corrected logical qubits in quantum annealing is unclear. An alternative approach to addressing qubit error is to use post-processing error mitigation [18-20]. Various error-mitigation techniques have been developed, but with limited success.

Recently, a quantum computing error mitigation method has been invented [21], and spin error mitigation for optimisation (SEMO) software [22] has been implemented to address this limitation. A potential solution-time advantage over classical methods has been demonstrated for segmenting 3D images using SEMO error-mitigated QA [23].

This study focuses on QA rather than QAOA, as a quantum annealer often has more qubits than a gate-model quantum computer. In the following sections, the weighted Max-Cut problem with mixed-signed edge weight values is solved using the standard D-Wave (DW) QA, the DW hybrid solver BQM, classical simulated annealing (SA), Tabu search (TS), and the GW-style algorithms. An error-mitigated QA algorithm will be introduced using the SEMO software. The accuracy and effectiveness of the SEMO error-mitigated QA solutions will be compared with those of the alternative approaches.

The QA computations in this study were performed with the DW Advantage and Advantage II quantum annealers [24]. All classical computations, including SEMO, classical SA, TS, GW and GW-inspired computations, were performed with a DELL OptiPlex 7070 desktop running MS-Windows 11. The computations were implemented in Python 3.11. The Python script and computational results are available for download from the CSIRO Data Access Portal [25].

## 2. Quantum annealing and classical computations

The solution of the weighted Max-Cut problem on a cubic lattice is formulated as computing the minimum value of the objective function

$$H = \sum_{<i,j>} J_{i,j}\, s_i\ s_j \tag{1}$$

where $s_i$ $(= \pm 1)$ is the discrete spin variable on the cubic lattice node $i$ $(i = 1,2,\cdots,N)$ and $N$ is the total number of nodes in the lattice. The lattice nodes are divided in two subsets. All nodes with spin value -1 belong to one subset and the other nodes with spin value +1 belong to the other subset. $J_{i,j}$ is the edge weight connecting spins $s_i$ and $s_j$, which takes a random value in the range $[w_0, 1]$. A global optimal solution to the above problem is a set of spin values for each lattice node such that $H$ takes the minimum value.

The QA solution of the weighted Max-Cut problem on a cubic lattice with $11^3$ nodes was performed in the DW Ocean Python environment [26]. The problem was embedded on the quantum annealers with the *minorminer* API. Larger lattices cannot be embedded due to the limited number of qubits.

Five batches of standard QA computations were performed for each value of $w_0 \geq -0.15$ using the DW Advantage and $w_0 \geq -0.25$ using the DW Advantage II quantum annealers. Fifty batches were computed for each value of $w_0 \leq -0.16$ with the DW Advantage and $w_0 \leq -0.26$ with the DW

Advantage II machines. Each batch includes $m = 2000$ reads with a read time of $\tau = 0.1$ ms. The total number of computations for each $w_0$ was $M = 5m$ or $M = 50m$.

The numerical average solution time $\bar{t}$ per global optimal solution was estimated as the total computation time $T$ divided by the number of reads $R$ that returned the global minimum value of the objective function.

$$\bar{t} = T/R \qquad \text{when } R > 0 \tag{2}$$

The numerical average QA computation time per global optimal solution $\bar{t}$ is shown in Figure 1. The QA computation time was based on the QPU access time. It does not include time to calculate the QPU embedding, which took up to 8 minutes on the DW Advantage and 6 minutes on the Advantage II machine. It also does not include time for QPU job queuing and cloud network delays. In addition to the QA computations, for each $w_0$, 5 batches of DW BQM computations were performed with $m = 10$ reads for each batch. For lattice size $11^3$, the DW BQM API requires a minimum solver time limit of $\tau = 3.7$ s. For $w_0 \geq -0.34$, BQM consistently returned global optimal solutions $R > 0$ for each computation. For more negative values of $w_0$, longer BQM time limits $\tau = 4.5$ s was used for $w_0 = -0.35$ and $\tau = 6$ s for $w_0 = -0.36$ and $-0.37$. The average time $\bar{t}$ per global optimal solution for the BQM computation results is also shown in Figure 1. The BQM computation time was based on the BQM API charge time. For $w_0 \geq -0.34$, the computation time is longer than the specified time limit, as some computations did not achieve the global optimal solution.

The numerical global optimal solution success probability $\bar{p}$ is estimated to be

$$\bar{p} = R/M \tag{3}$$

Assuming that the global optimal solution follows a binomial distribution, the confidence interval is estimated as [27],

$$p = \bar{p} \pm \Delta p = \bar{p} \pm zM \tag{4}$$

For a $p_c = 95\%$ confidence level, $z = 1.96$. $p_L = \bar{p} - \Delta p$ is the lower bound probability and $p_U = \bar{p} + \Delta p$ is the upper bound probability. The average time for the global optimal solution lower bound $t_L$ and upper bound $t_U$ were estimated as

$$\begin{cases} t_L = T/Mp_U \\ t_U = T/Mp_L \end{cases} \tag{5}$$

Equation (5) is used to estimate the solution time error bars in Figure 1 and subsequent figures.

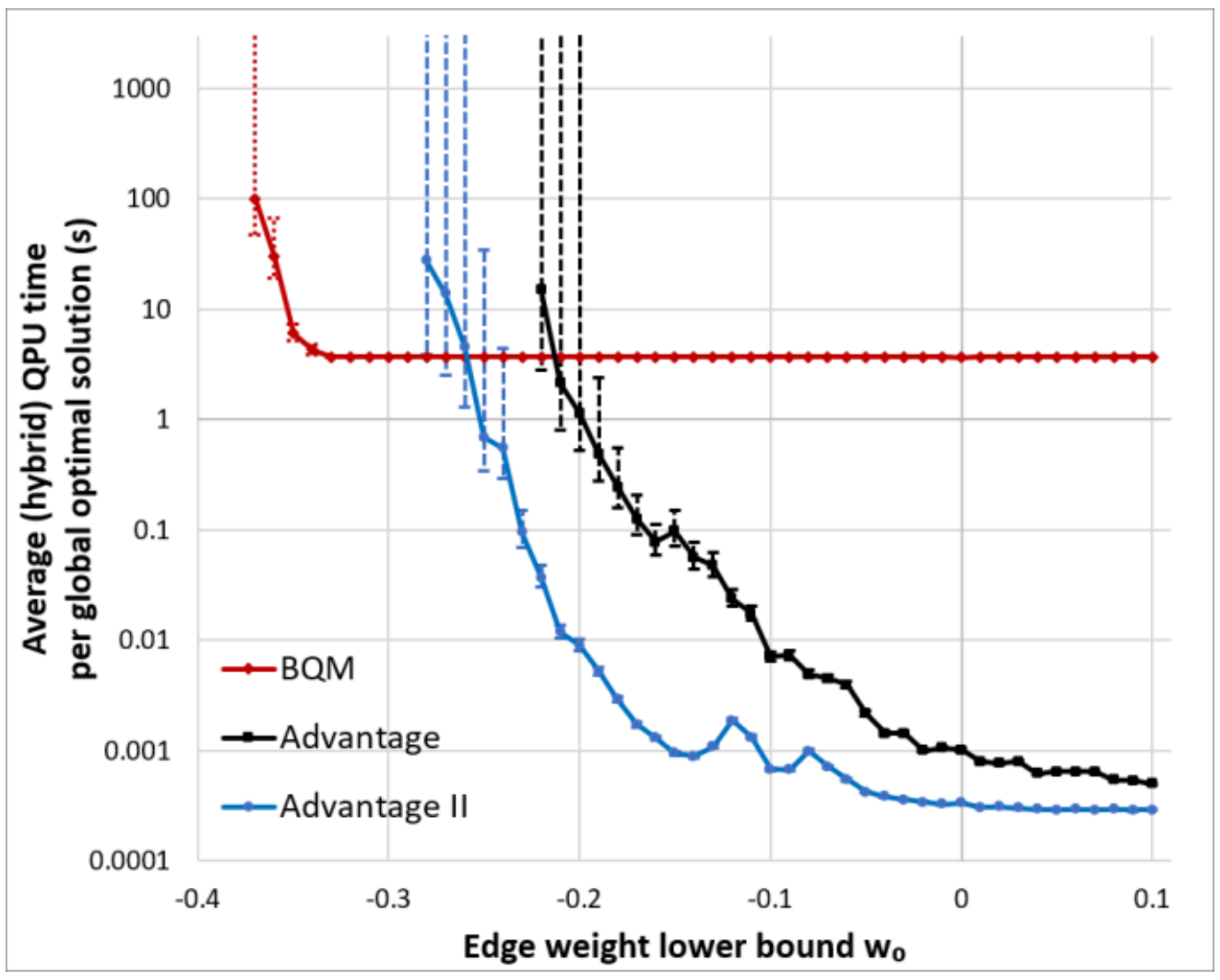

**Figure 1.** DW QA and BQM average solution time $\bar{t}$ per global optimal solution versus $w_0$ on a cubic lattice with $11^3$ nodes. The error bars are based on a binomial distribution of global optimal solution success probability.

Figure 1 shows that a quantum annealer can efficiently solve the weighted Max-Cut problem within the order of milliseconds on a $11^3$ cubic lattice for non-negative edge weights. As the lower bound of the edge weights becomes more negative, the QA solution time increases exponentially. The computation indicated that up to 100 000 reads, no reliable global optimal solution was obtained for $w_0 < -0.22$ and $w_0 < -0.28$ for the Advantage and Advantage II machines, respectively. Figure 1 also shows that the DW Advantage II quantum annealer is an order of magnitude more efficient than the Advantage machine. Compared with the Advantage II machine, the BQM solver showed some solution time advantage near the negative end of $w_0$, although it was inefficient by orders of magnitude toward the high-value end of $w_0$. At $w_0 = -0.37$, the effective time $\bar{t}$ per global optimal solution is about 100 s, which is likely to be too long for certain practical applications.

The weighted Max-Cut problem in Equation (1) can be solved with appropriate classical solvers. The DW Ocean simulated annealing (SA) and Tabu search (TS) solvers were used for classical computational solutions. For SA, the computational efficiency is sensitive to the number of sweeps. For the $11^3$ cubic lattice and $w_0 = -0.35$, the average SA solution time per global optimal solution $\bar{t}$ is shown in Figure 2, which reaches a minimum at approximately 10,000 sweeps. The optimal number of sweeps, which is 10,000, was selected for all SA computations in this study.

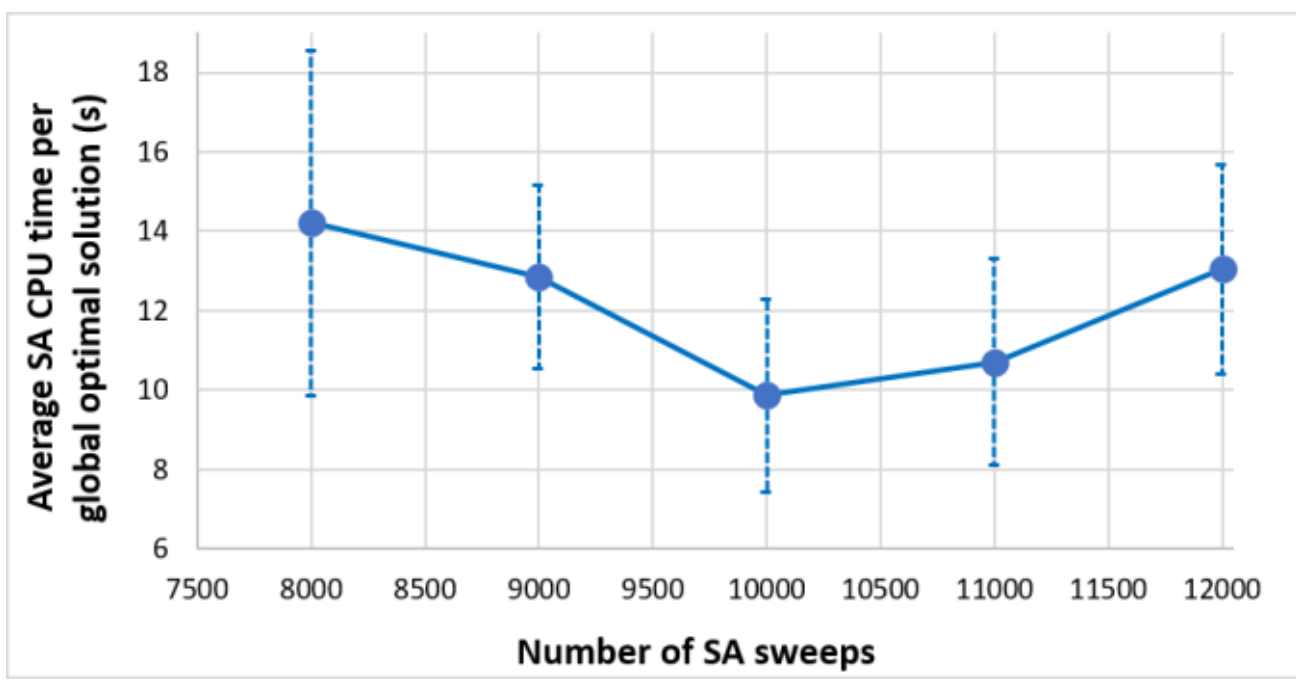

**Figure 2.** SA average solution time $\bar{t}$ per global optimal solution versus the number of sweeps on a cubic lattice with $11^3$ nodes and $w_0 = -0.35$. The error bars are based on the sample standard deviation $\pm\sigma$ between batches.

Five batches of computations were performed for both SA and TS. The SA inverse temperature $\beta$ range was optimised by the DW Ocean platform. The tenure value for TS was set to the Ocean TS maximum value of 20. The classical SA and TS computations were performed on a DELL OptiPlex7070 desktop computer running MS Windows 11. The solution CPU times for SA and TS are shown in Figure 3. The SA and TS computation CPU times were based on the "wall-clock time" for calling the respective SA and TS API functions.

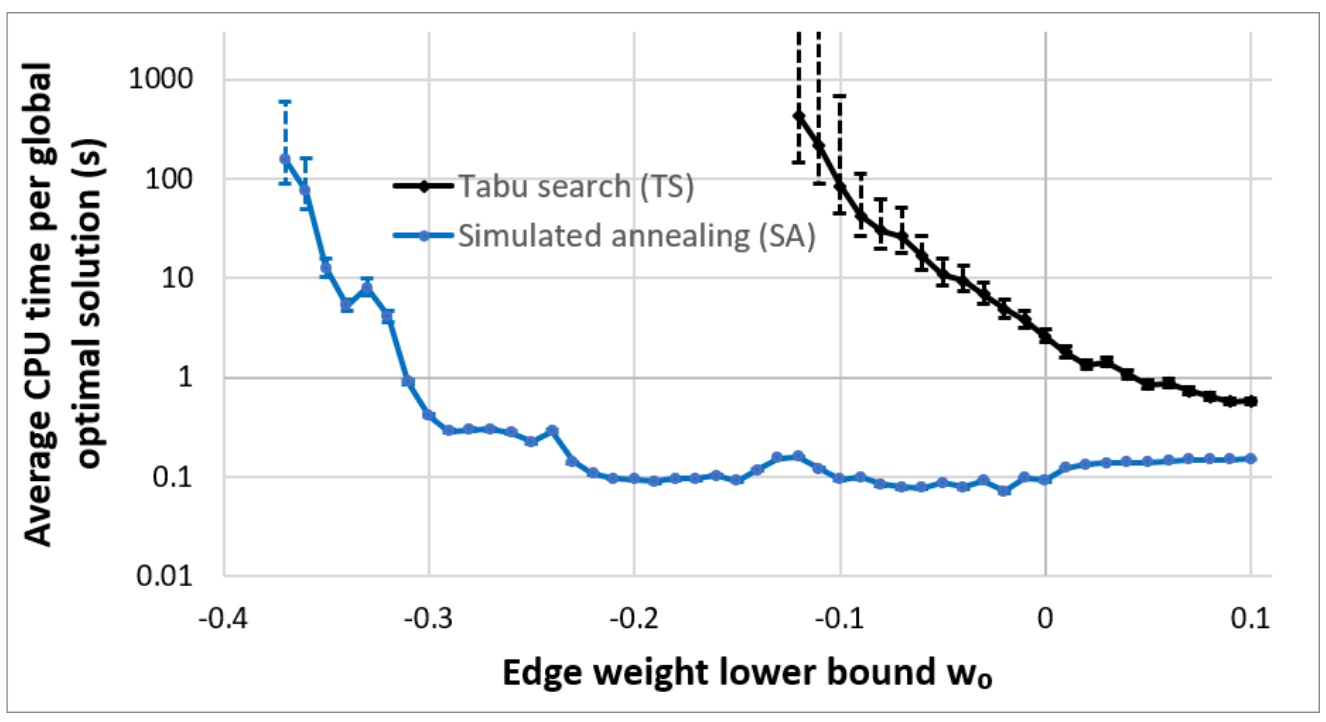

**Figure 3.** Classical SA and TS average solution time $\bar{t}$ per global optimal solution versus $w_0$ on a cubic lattice with $11^3$ nodes. The error bars are based on the sample standard deviation $\pm\sigma$ between batches.

Comparing solution time with the techniques above, Figures 1 and 3 indicate that the standard QA with the DW Advantage II quantum annealer is the most efficient method for $w_0 > -0.24$ followed by the DW Advantage for $w_0 > -0.16$. For further lower values of $w_0$ down to $w_0 = -0.34$, SA is the most efficient choice. BQM becomes the most efficient method for $w_0 < -0.34$. TS is slower than all the other methods.

## 3. Error-mitigated quantum annealing

The deteriorating performance of the QAs as $w_0$ decreases may be related to qubit errors in the quantum annealer QPUs. All embedded qubits in a quantum annealer's QPU must perform correctly for the processor to produce a correct answer. However, a physical qubit in a quantum processor may deviate from its ideal state due to factors such as environmental noise. This leads to an exponential decrease in the probability that all qubits simultaneously collapse to the correct measured values upon completion of the computation [23, 28]. In this section, a SEMO error-mitigated quantum annealing algorithm [25] for Max-Cut will be introduced to mitigate the negative impact of the qubit error. SEMO is an implementation of a patent-pending method [21] for systematically adjusting the state of coupled clusters of variables to minimize the objective function value.

For each value of $w_0$ on a cubic lattice with $11^3$ nodes, the Max-Cut problem was solved by incorporating the SEMO APIs with the following algorithmic procedures:

(1) Generate random edge weights distributed in $[w_0, 1]$ between neighbouring nodes on the lattice.
(2) Pass the edge weight values to the SEMO module using the API `SetupIsing`.
(3) Embed the Max-Cut edge weight coefficients on the quantum annealer.
(4) Perform quantum annealing computation with 2000 reads for each batch of computation.
(5) Perform SQC equivalent SEMO (1,0) computation of QA results, with SEMO API parameter setting `SetErrMitMethod(1,0)`.
(6) Sort the solutions with the objective function value in ascending order.
(7) Compute dot products for consecutive pairs of solutions. For each pair, if the dot product is negative, all spins in the second solution of the pair are multiplied by –1.
(8) Allowing up to 50 inconsistencies for each spin using the SEMO API `SetDataTolerance(-1,50)`.
(9) Perform SEMO (2,2) computation with SEMO API parameter setting `SetErrMitMethod(2,2)`.
(10) Repeat 5 times (or 50 times for 50 batches) from (3) or (4) and return solutions with the minimum objective function value.

In step (7), the operation takes advantage of the symmetry property of the Max-Cut problem that multiplying all spins by –1 does not change the objective function value. That is, the objective function as defined in Equation (1) is invariant under global spin reversal. The sample alignment is necessary for better SEMO computational efficiency.

In step (8) for Advantage II and out of 2000 solutions for each batch, subsets of 1000 and 200 low objective function value solutions were processed for $-0.35 \geq w_0 \geq -0.32$ and $w_0 < -0.35$, respectively, to improve the SEMO computational efficiency further. For these subsets, the error tolerance setting was `SetDataTolerance(-1)`.

The solution time for SEMO error-mitigated QA is shown in Figure 4. The solution time was based on the sum of the QPU access time and the "wall clock time" CPU time for SQC and SEMO computations. Similar to Figure 1, the solution time does not include the QPU embedding time, QPU job queuing time or cloud network delays. For convenience of comparisons, Figure 4 also includes the average solution times for standard QA, BQM, SA and TS.

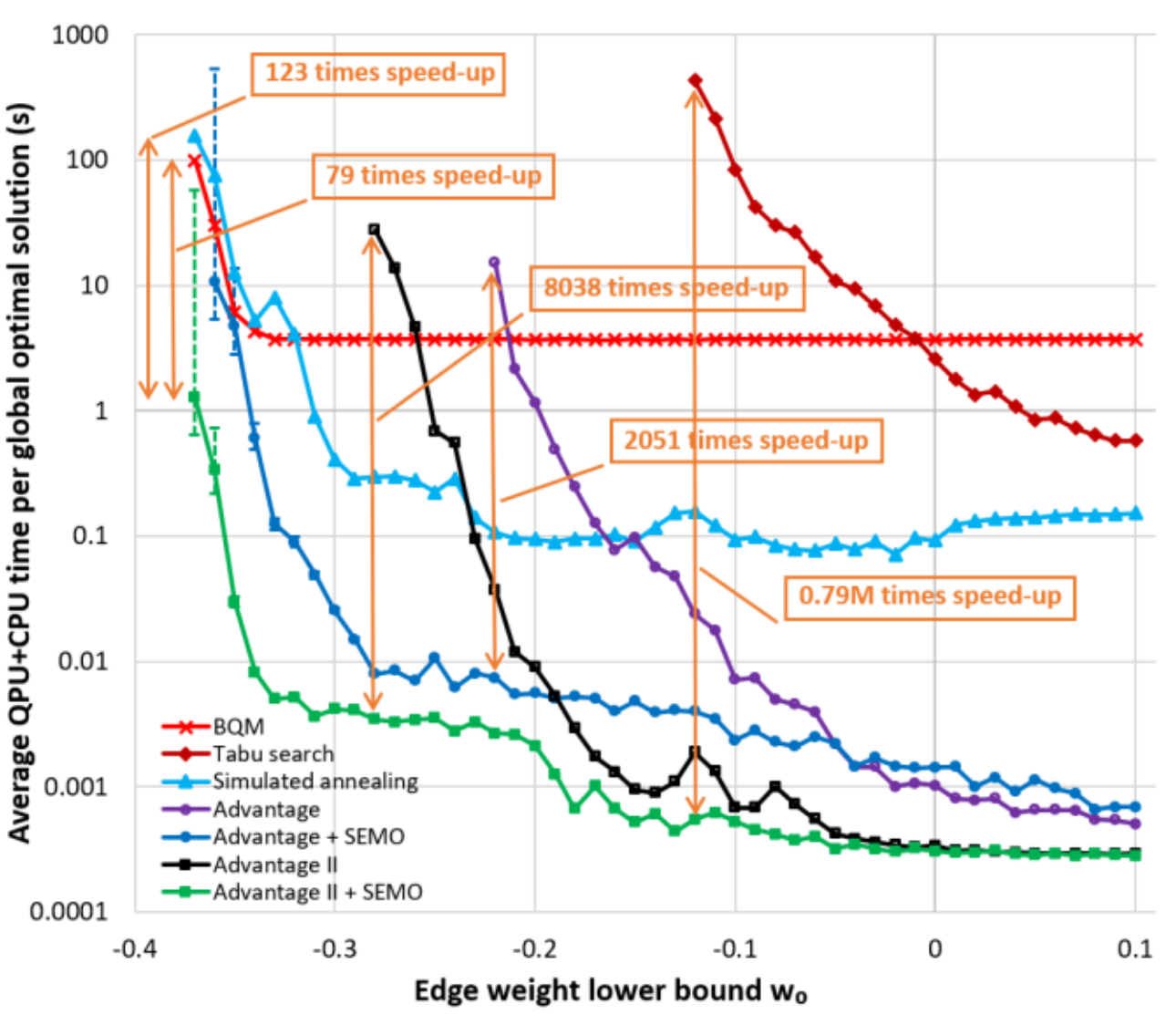

**Figure 4.** SEMO error-mitigated QA average solution time $\bar{t}$ per global optimal solution versus $w_0$ on a cubic lattice with $11^3$ nodes. The error bars are based on the binomial distribution of global optimal solution success probability. The SEMO computation includes SEMO method (1,0) followed by the SEMO method (2,2). Average solution times for other methods are included for comparison.

In Figures 1 and 5 for $w_0 = -0.22$, the average solution time per global optimal solution with the DW Advantage standard QA was 15 s, while the solution time with SEMO error-mitigated QA was 7.3 ms. That is, the SEMO error-mitigated QA delivered a speed-up factor of over 2000. For $w_0 = -0.28$, the average solution time per global optimal solution using the DW Advantage II standard QA was 28 s, whereas the solution time with SEMO error-mitigated QA was 3.4 ms. That is, the SEMO error-mitigated QA delivered a speed-up factor of over 8000. This demonstrates that SEMO error-mitigated QA is significantly more efficient at obtaining global optimal solutions than standard QA. The improvement increases as the problem becomes more complex; that is, when $w_0$ becomes more negative. The comparison also showed that the SEMO error-mitigated QA with DW Advantage II is faster than the DW BQM solver. At $w_0 = -0.37$, the global optimal solution time for BQM is 100 s and that for the SEMO error-mitigated annealing with Advantage II is 1.27 s, which gives a solution time ratio of about 80.

At $w_0 = -0.37$, the average time per global optimal solution for SA is 156 s. That is, the SEMO error-mitigated QA with DW Advantage II is over 100 times faster than classical SA. The classical TS did not reliably produce a global optimal solution for $w_0 < -0.12$. At $w_0 = -0.12$, the average time per global optimal solution with TS is 432 s while it is 0.54 ms with SEMO error-mitigated QA on DW Advantage II. The time difference ratio is about 0.8 million. The numerical comparisons demonstrated that SEMO error-mitigated QA is consistently faster than other solvers for Max-Cut problems with mixed-sign weights.

As the value of $w_0$ becomes more negative in Figure 4, the solution time for all methods shows an exponentially increasing trend. At the same time, the number of computations that returned global optimal solutions decreases, as shown in Figure 5. That is, the Max-Cut problem on a cubic lattice becomes more challenging as the edge-weight lower bound becomes more negative.

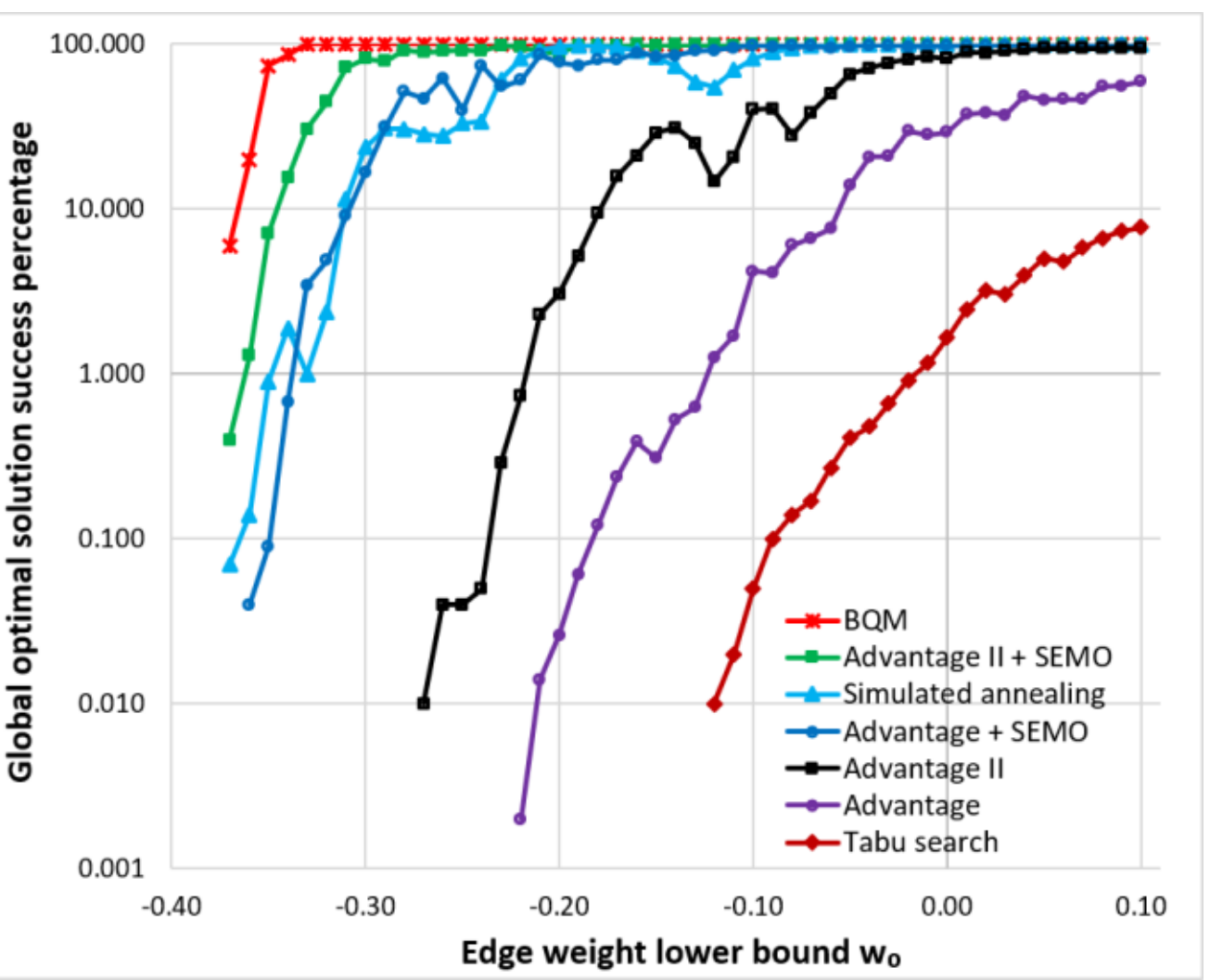

**Figure 5.** Percentages of computational reads that returned global optimal solutions versus $w_0$ on a cubic lattice with $11^3$ nodes.

Figure 5 shows that DW BQM has the highest percentage of computations that produce a global optimal solution, followed by the SEMO error-mitigated QA with Advantage II. Again, TS performs the worst. For a $p_c = 95\%$ confidence, the expected number of computations $G$ to obtain a global optimal solution can be estimated as

$$G = \max\left[1, \frac{\ln(1-p_c)}{\ln(1-R/M)}\right] \quad \text{when } R > 0 \qquad (6)$$

The dependence of the expected number of computations $G$ on the edge lower bound $w_0$ is shown in Figure 6. Again, the DW BQM requires the smallest number of computations to reach a global optimal solution. Figure 6 illustrates that the SEMO error-mitigated QA demonstrated a significant reduction in the required number of computations as compared with the standard QA.

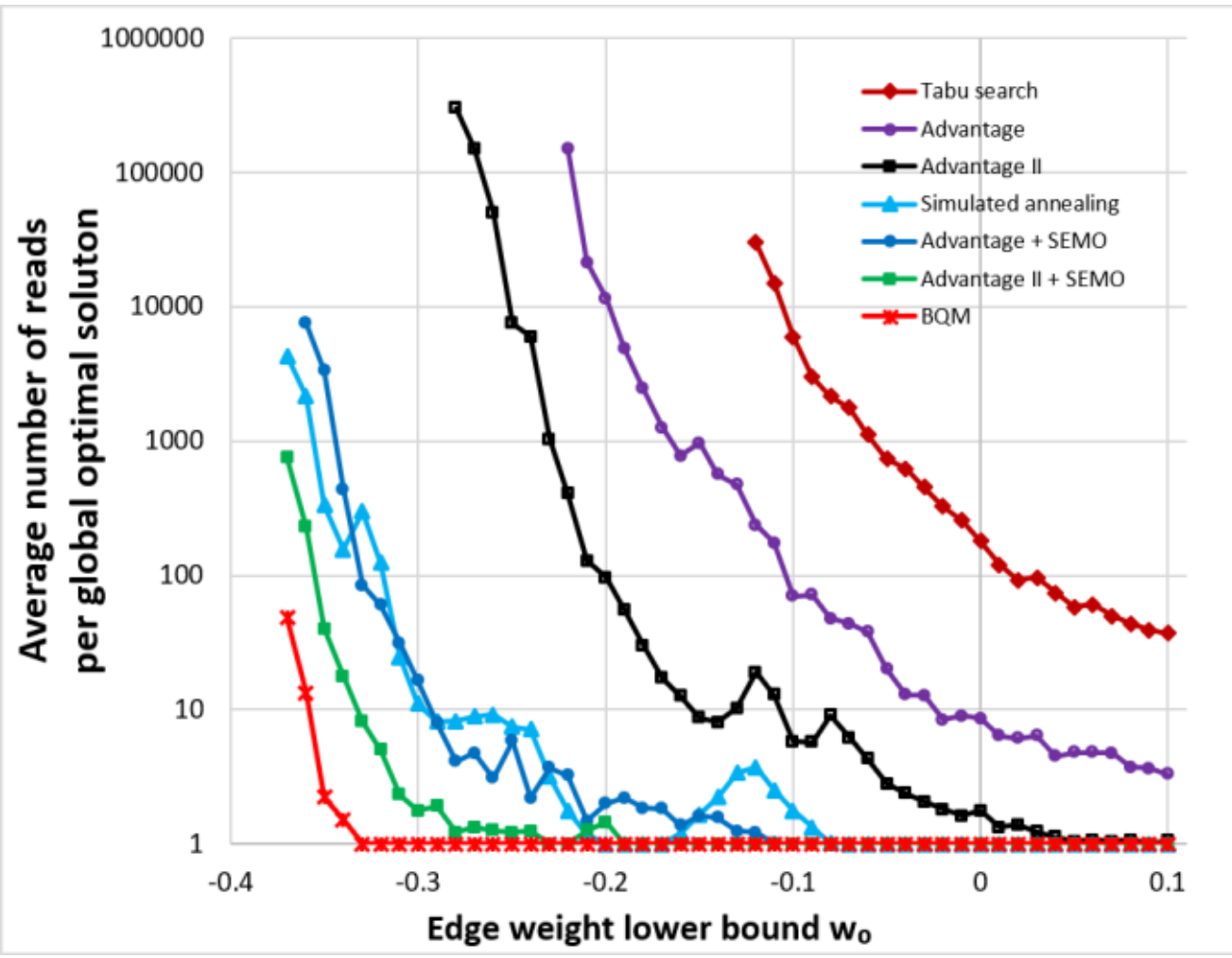

**Figure 6.** Expected number of computations per global optimal solution versus $w_0$ on a cubic lattice with $11^3$ nodes.

For each value of $w_0$, the putative global optimal objective function value is defined as the minimum objective function value obtained by all computational methods used in this study, including the higher order SEMO error-mitigation methods SEMO (2,3) and SEMO (2,4), which were achieved with parameter settings `SetErrMitmethod(2,3)` and `SetErrMitMethod(2,4)`. These global optimal (minimum) objective function values are listed in Table 1. With SEMO (2,2), the SEMO error-mitigated QA consistently produced the global minimum objective function value for $-0.37 \leq w_0 \leq 0.10$.

**Table 1.** Global minimum objective function and max edge cut values on a cubic lattice with $11^3$ nodes and random neighbouring edge weights for $-0.37 \geq w_0 \geq 0.1$.

| w0 | Min obj func | Max edge cut | GW edge cut |
|---|---|---|---|
| 0.10 | -2007.4619 | 4014.9238 | 4014.9238 |
| 0.09 | -1989.4337 | 3978.8674 | 3978.8674 |
| 0.08 | -1971.4055 | 3942.8110 | 3942.8110 |
| 0.07 | -1953.3773 | 3906.7546 | 3906.7546 |
| 0.06 | -1935.3491 | 3870.6982 | 3870.6982 |
| 0.05 | -1917.3209 | 3834.6418 | 3834.6418 |
| 0.04 | -1899.2927 | 3798.5854 | 3798.5854 |
| 0.03 | -1881.2645 | 3762.5290 | 3762.5290 |
| 0.02 | -1863.2363 | 3726.4726 | 3726.4726 |
| 0.01 | -1845.2081 | 3690.4162 | 3690.4162 |
| 0.00 | -1827.1799 | 3654.3598 | 3654.3598 |
| -0.01 | -1809.1517 | 3618.3034 | 3618.3034 |
| -0.02 | -1791.1235 | 3582.2470 | 3582.2470 |
| -0.03 | -1773.0953 | 3546.1906 | 3546.1906 |
| -0.04 | -1755.0671 | 3510.1342 | 3510.1342 |
| -0.05 | -1737.0389 | 3474.0778 | 3474.0778 |
| -0.06 | -1719.0107 | 3438.0214 | 3438.0214 |
| -0.07 | -1700.9825 | 3401.9650 | 3401.9650 |
| -0.08 | -1682.9803 | 3365.9346 | *3365.9086* |
| -0.09 | -1665.0634 | 3329.9895 | *3329.8522* |
| -0.10 | -1647.1466 | 3294.0445 | *3293.7958* |
| -0.11 | -1629.2297 | 3258.0994 | *3257.7394* |
| -0.12 | -1611.3539 | 3222.1954 | *3215.0355* |
| -0.13 | -1593.6358 | 3186.4491 | *3178.9912* |
| -0.14 | -1575.9689 | 3150.7540 | *3144.6027* |
| -0.15 | -1558.3019 | 3115.0588 | *3108.5553* |
| -0.16 | -1540.6350 | 3079.3637 | *3072.5080* |
| -0.17 | -1522.9681 | 3043.6685 | *3036.4607* |
| -0.18 | -1505.3581 | 3008.0304 | *2998.7696* |
| -0.19 | -1487.7934 | 2972.4375 | *2962.7252* |
| -0.20 | -1470.2647 | 2936.8806 | *2916.2672* |
| -0.21 | -1452.8099 | 2901.3975 | *2880.9790* |
| -0.22 | -1435.4111 | 2865.9706 | *2846.6353* |
| -0.23 | -1418.0620 | 2830.5933 | *2811.3968* |
| -0.24 | -1400.7311 | 2795.2342 | *2770.9965* |
| -0.25 | -1383.6614 | 2760.1363 | *2732.9877* |
| -0.26 | -1366.7406 | 2725.1873 | *2688.7208* |
| -0.27 | -1350.1200 | 2690.5384 | *2648.2384* |
| -0.28 | -1333.8239 | 2656.2142 | *2602.0584* |
| -0.29 | -1317.6678 | 2622.0299 | *2557.4037* |
| -0.30 | -1301.6669 | 2588.0007 | *2489.2922* |
| -0.31 | -1285.7105 | 2554.0161 | *2446.2078* |
| -0.32 | -1269.7610 | 2520.0385 | *2375.5395* |
| -0.33 | -1254.1255 | 2486.3747 | *2335.5974* |
| -0.34 | -1239.1010 | 2453.3220 | *2301.4635* |
| -0.35 | -1224.4381 | 2420.6309 | *2255.0322* |
| -0.36 | -1210.1435 | 2388.3081 | *2206.8619* |
| -0.37 | -1196.3226 | 2356.4591 | *2139.5970* |

For each value of $w_0$, the maximum edge cut $W$ is calculated as

$$W = \sum_{<i,j>} w_{i,j} \tag{7a}$$

where

$$w_{i,j} = \begin{cases} -J_{i,j} & \text{when } i \neq j \\ 0 & \text{otherwise} \end{cases} \tag{7b}$$

The max edge cut values $W$ calculated from Equation (7) are also listed in Table 1. Numerically, the minimum objective function value and the max edge cut value have a one-to-one relationship. That is, when the objective function in Equation (1) takes the minimum value, the Max-Cut in Equation (7) takes the maximum value.

The edge-weight matrix on a cubic lattice is sparse. This makes the original GW algorithm inefficient when the number of nodes is large. The accuracy of the GW-style algorithms may also be limited when $w_0$ is negative. In certain cases, a quantum-inspired algorithm showed improved computational efficiency over GW-style algorithms [29]. For large sparse graphs in this paper, we approximate the computation using a spectral algorithm inspired by the GW algorithm [25, 30]. Such a GW-inspired algorithm is computationally efficient. It takes about $30 \pm 5$ ms to compute for a $11^3$ lattice. In comparison, the same computation took 6.5 hours with the original GW algorithm. As shown in Table 1, for $w_0 \geq -0.07$, the GW-inspired algorithm produced the same optimal max cut. However, for $w_0 \leq -0.08$, the GW-inspired algorithm did not reach the optimal max cut.

## 4. Summary and outlook

The weighted Max-Cut problem was investigated on a cubic lattice with $11^3$ nodes and random neighbouring weights. The lattice is selected as it has the largest number of nodes that can be embedded on the DW quantum annealers. Without loss of generality, the upper bound on the weight was fixed at 1, and various lower bounds were investigated using several computational techniques. These techniques include standard D-Wave quantum annealing with the Advantage and Advantage II quantum annealers, D-Wave BQM hybrid solver, classical simulated annealing, Tabu search, and Goemans–Williamson-style solvers. A SEMO error-mitigated quantum annealing algorithm was introduced.

When the edge weights were non-negative, the problem could be solved within a reasonable time with various classical and quantum annealing solvers. The most efficient technique was D-Wave quantum annealing, with a sub-millisecond solution time, and the most inefficient was the D-Wave BQM solver, which required a minimum solver time of 3.7 s. SEMO error-mitigated quantum annealing was slightly less efficient than standard quantum annealing due to the additional computational overhead. Classical simulated annealing, Tabu search and the Goemans–Williamson-inspired solver required a longer computing time to obtain a global optimal solution than standard quantum annealing.

As the lower bound of the edge weight became more negative, the Goemans–Williamson-inspired solver failed to produce global optimal solutions. For Tabu search and standard quantum annealing, the efficiency of obtaining a global optimal solution dropped exponentially and became impractical for moderately negative edge weights. Simulated annealing, BQM and SEMO error-mitigated quantum annealing continued to produce global optimal solutions until the edge weight reached –0.37, beyond which they became impractical. SEMO error-mitigated quantum annealing gave an exponential improvement in solution time over standard quantum annealing. SEMO error-mitigated quantum annealing with D-Wave Advantage II was significantly more efficient than the other techniques in computing the global optimal solution. That is, by mitigating qubit errors in quantum annealing computational results, SEMO effectively extended the capability for quantum annealers to solve such complex optimisation problems efficiently.

For larger lattices with more than $11^3$ nodes, standard and error-mitigated quantum annealing are no longer applicable due to the limited number of qubits in the commercially available quantum annealer QPUs. D-Wave BQM and classical simulated annealing remain suitable, and BQM would have a degree of solution time advantage over simulated annealing. Integration of SEMO with BQM or simulated annealing may further improve efficiency. For non-negative edge weights, the Goemans–Williamson-style algorithms would also be suitable.

The investigation indicated that the optimal choice of solvers depends on factors such as the edge weights and problem size. It also depends on the priority requirements. If solution time is the primary focus and the problem can be embedded in the available quantum annealing hardware, SEMO error-mitigated quantum annealing would be the optimal choice. If the number of computations is the prime consideration, BQM would be the best option. Given the high cost of quantum computation, classical simulated annealing should be considered if cost is the limiting factor.

It should be noted that, although SEMO error-mitigated annealing outperformed other solvers for Max-Cut with mixed-sign edge weights, the results should not be interpreted as a demonstration of quantum annealing advantage, as we cannot rule out the existence of other, more efficient classical solvers. The time required to calculate the QPU embedding is an important factor affecting the overall computational efficiency of standard and SEMO error-mitigated quantum annealing. Further advances, including more efficient QPU embedding algorithms and the planned release of DW Advantage III with 100,000 qubits [31], will broaden the scope of SEMO error-mitigated quantum annealing for solving larger combinatorial optimisation problems. The time-efficient solution capability of SEMO error-mitigated quantum annealing in solving complex combinatorial optimisation problems will be particularly impactful for time-critical applications such as real-time management of major sporting events, traffic control in busy districts, energy and communication network management, and certain defence scenarios.

## Acknowledgements

The authors would like to acknowledge useful discussions with Clement Chu and Krzysztof Giergiel of CSIRO, Michael Hall of NEC Australia, Pau Farre, Andrew King, and Catherine Potts of D-Wave, and Vu Ngo and Kien Nguyen of Queensland University of Technology. Support by CSIRO's Quantum Technology Future Science Platform is also acknowledged.

Data Access Statement: Research data supporting this publication are available at reference [25]: *CSIRO Data Collection*. https://doi.org/10.25919/vrqd-tw70 or PID: https://data.csiro.au/collection/csiro:73984.

Conflict of Interest Declaration: The authors are co-inventors, and CSIRO is the owner, of the pending patent [21] on which the SEMO method is based.